\documentclass[a4paper,fleqn,usenatbib, useAMS]{mnras}
\usepackage{graphicx}
\usepackage{multicol}
\usepackage{enumerate}
\usepackage{placeins}
\usepackage{amsmath}
\usepackage{pdflscape}
\usepackage{amssymb}
\usepackage{longtable}

\usepackage{journal_names}

\def\gaia{{\it Gaia }}
\title[Fast \gaia transients] {The fast transient sky with \gaia}
\author[Wevers et al.]{Thomas Wevers$^{1}$\thanks{Email: t.wevers@astro.ru.nl}, Peter G. Jonker$^{2,1}$, Simon T. Hodgkin$^{3}$, \newauthor Zuzanna Kostrzewa-Rutkowska$^{2,1}$, Diana L. Harrison$^{3,4}$, Guy Rixon$^{3}$, Gijs Nelemans$^{1,5}$, \newauthor Maroussia Roelens$^{6}$, Laurent Eyer$^{6}$, Floor van Leeuwen$^{3}$ and Abdullah Yoldas$^{3}$ \\\\
$^{1}$Department of Astrophysics/IMAPP, Radboud University, P.O. Box 9010, NL-6500GL Nijmegen, The Netherlands\\
$^{2}$SRON, Netherlands Institute for Space Research, Sorbonnelaan 2, NL-3584CA Utrecht, The Netherlands\\
$^{3}$Institute of Astronomy, Madingley Road, Cambridge CB3 0HA, United Kingdom\\
$^{4}$Kavli Institute for Cosmology, University of Cambridge, Madingley Road, Cambride CB3 0HA, United Kingdom\\
$^{5}$Institute for Astronomy, KU Leuven, Celestijnenlaan 200D, B-3001 Leuven, Belgium\\
$^{6}$Department of Astronomy, University of Geneva, Ch. des Maillettes 51, CH-1290 Versoix, Switzerland\\
}

\begin{document}
\date{}
\pagerange{\pageref{firstpage}--\pageref{lastpage}} \pubyear{2017}
\maketitle
\label{firstpage}

\begin{abstract}
The ESA \gaia satellite scans the whole sky with a temporal sampling ranging from seconds and hours to months. Each time a source passes within the \gaia field of view, it moves over 10 CCDs in 45 s and a lightcurve with 4.5 s sampling (the crossing time per CCD) is registered. Given that the 4.5 s sampling represents a virtually unexplored parameter space in optical time domain astronomy, this data set potentially provides a unique opportunity to open up the fast transient sky. We present a method to start mining the wealth of information in the per CCD \gaia data. We perform extensive data filtering to eliminate known on-board and data processing artefacts, and present a statistical method to identify sources that show transient brightness variations on $\lesssim$\,2 hours timescales. We illustrate that by using the \gaia photometric CCD measurements, we can detect transient brightness variations down to an amplitude of 0.3 mag on timescales ranging from 15 seconds to several hours. We search an area of $\sim$\,23.5 square degrees on the sky, and find four strong candidate fast transients. Two candidates are tentatively classified as flares on M-dwarf stars, while one is probably a flare on a giant star and one potentially a flare on a solar type star. These classifications are based on archival data and the timescales involved. We argue that the method presented here can be added to the existing \gaia Science Alerts infrastructure for the near real-time public dissemination of fast transient events.
\end{abstract}


\section{Introduction}
During the last decade, a number of intriguing new classes of short timescale transient events have been discovered. Such discoveries often happen serendipitously or after analysis of archival data, implying that the transient has faded to pre-detection levels long before any follow up observations can be performed. These discoveries can trigger a suite of new surveys and experiments in an attempt to uncover and characterize the new classes of transients. For example, the detection of a bright milli-second duration burst in archival radio data by \citet{Lorimer2007} prompted a new field of radio astronomy devoted to these so-called fast radio bursts (FRBs). Despite the subsequent discovery of additional events, including a repeating FRB (e.g. \citealt{Champion2016, Spitler2016}), the nature of these short, probably extragalactic flashes remains a mystery with many theories to explain its properties (\citealt{Chatterjee2017}, and references there-in). In another part of the electromagnetic spectrum, several fast ($\sim$\,10s\,--\,100s of seconds duration) large amplitude X-ray flashes were found (e.g. \citealt{Jonker2013, Glennie2015, Bauer2017}) in archival X-ray observations. Similarly to the fast radio bursts, the nature of these events is still debated. 

At optical wavelengths, there are several classes of known and theoretically predicted short duration transients ranging from flares on M-dwarf and solar-type stars \citep{Maehara2015} to the prompt optical emission of gamma-ray bursts \citep{Yi2017} and the tidal disruption of white dwarfs (WDs) by an intermediate mass black holes (IMBHs; \citealt{Rosswog2009,Macleod2016}). A disadvantage of optical observations compared to X-ray or radio observations is that it is much harder to obtain deep, high time resolution observations of a significant area on the sky. As a consequence, the parameter space of optical variability on $\sim$\,second timescales is still virtually unexplored. Early work by e.g. \citet{Pedersen1984} was not followed up spectroscopically. Current optical sky surveys, such as the All-Sky Automated Survey for Supernovae (ASAS--SN), intermediate Palomar Transient Factory (iPTF) or the Panoramic Survey Telescope and Rapid Response System (Pan-STARRS; \citealt{Chambers2016}), are sampling the sky with cadences down to a few hours (see e.g. \citealt{Berger2013}), while the upcoming Large Synoptic Survey Telescope (LSST) will reach a cadence of $\sim$\,20 minutes \citep{lsstsciencebook}. There exist some surveys that reach cadences of one to a few minutes (e.g. the OmegaWhite Survey, \citealt{Macfarlane2015} or the Kepler K2 short cadence mission e.g. \citealt{Gilliland2010}), but survey depth or sky coverage are usually sacrificed  to reach shorter cadences.

The \gaia satellite \citep{Prusti2016, Brown2016} is performing an all sky survey with a range of cadences going from seconds to hours and weeks, and therefore has the potential to be used to search for variable stars \citep{Clementini2016, Eyer2017}, traditional transients \citep{Hodgkin2013} and fast variables \citep{Roelens2017} and transients. Ultimately, it will provide lightcurves at a minimum time resolution of 4.5 s for all sources brighter than $G$\,=\,20.7 mag. Therefore \gaia will provide a unique data set including the first all sky survey with $\sim$\,seconds time resolution at optical wavelengths. Given that this parameter space is still largely unexplored, \gaia has the potential to revolutionize our view of the fast transient optical sky.

Software infrastructure to exploit the \gaia all sky data to search for and disseminate alerts of transient events in near real-time already exists \citep{Hodgkin2013}. The data processing pipeline and search algorithms will be described in detail in Hodgkin et al. (in prep.). The data processing and filtering employed in our work is similar in scope and nature, with the important difference that the current \gaia Alerts pipeline uses median source magnitudes per transit (i.e. the median of the 9 astrometric field CCD measurements) as the input to their detection algorithms. In combination with the requirement of source detection as an outlier in two transits, this implies that such a search is not sensitive to transient events below 106 minutes but tailored to longer timescales such as outbursts in cataclysmic variables, (super)novae, tidal disruption events and microlensing events. The goal of this work is to explore the feasibility of using \gaia data to find transient events on short ($\leq$\,2 hours) timescales using the 4.5 s cadence lightcurves. 

The article is organized as follows: we describe the data and the filtering process to obtain a clean sample of \gaia data in Section \ref{sec:description}. Next, we describe our outlier detection algorithm in Section \ref{sec:algorithm}, and apply it to the \gaia data to illustrate the potential of \gaia as a fast transient detection machine in Section \ref{sec:fasttransientresults}. We summarize and discuss future developments in Section \ref{sec:summaryandfuture}.

\section{Description of the Gaia data}
\label{sec:description}
The \gaia satellite is an ESA cornerstone mission that was launched in 2013 with the aim of mapping the brightnesses and positions of a billion stars \citep{Prusti2016}. Recently, the first \gaia data were released in Gaia Data Release 1 (GDR1; \citealt{Brown2016}). This data set consists of astrometry and photometry of over 1 billion stars brighter than $G$\,=\,20.7 mag. \gaia has currently observed a staggering 81 {\it billion} transits in its more than 1000 days of routine operations.

{\it Gaia} scans the sky with two telescopes pointing in directions separated by an angle of 106.5$^{\circ}$ on the sky. The satellite spins with a rate of 60 arcsec s$^{-1}$, leading to one full revolution of the satellite spin axis every 6 hours. A slow precession of the spin axis with a period of 63 days around the solar direction ensures full sky coverage (see \citealt{Lindegren2016} and \citealt{Leeuwen2017} for more details about the scanning law and sky coverage). The spinning motion of \gaia requires the detectors to be operated in time-delayed integration (TDI) mode, in which the charge accumulated due to an astrophysical source is propogated in the along-scan (AL) direction at a matching rate. Brightness measurements are available after each CCD crossing. Due to the scanning pattern, \gaia will observe each position on the sky on average between 50 and 250 times during the 5 year nominal mission duration, depending mainly on the ecliptic latitude. 

The two fields-of-view (FoVs) are projected on the same focal plane, which consists of 106 CCDs. The Gaia detector plane is described in detail in \citet{Leeuwen2017}; here we briefly introduce the most relevant concepts for our work. Two columns of 7 CCDs (called star mappers, or SMs) are used to perform the on-board source detection; each SM is associated with 1 FoV only, while the light from the other FoV is blocked by a baffle. The astrometric portion of the detector plane consists of 9 columns of 7 CCDs each. All these CCDs operate in the \gaia white-light band-pass (the $G$ filter; \citealt{Carrasco2016}). Verification of the source detection is performed on the first column of astrometric field (AF) CCDs. Each detected source is allocated a window\footnote{A window constitutes a number of pixels in the along-scan and across-scan directions.} (see e.g. Figure \ref{fig:parasitic}), the size of which depends mainly on the brightness. In addition each window has an associated gate and window class, again depending mainly on the brightness of the source. The allocated windows are propagated across the detector plane as the satellite scans the sky. Given the \gaia spin rate, it takes approximately 4.5 seconds for a source to traverse 1 CCD and $\sim$\,45\,--\,50\footnote{Sources detected on SM1 must traverse the physical size of SM2 before reaching AF1, leading to a lightcurve that spans 50 s instead of 45 s, although the number of measurements is the same for SM1 and SM2.} seconds to traverse the SM\,+\,AF portion of the focal plane. A source brightness measurement is available after each CCD crossing, yielding a lightcurve with a total span of 45\,--\,50 seconds and a cadence of 4.5 s. \gaia will thus provide (sparsely sampled) high time resolution lightcurves of the whole sky, with on average $\sim$\,500\,--\,2000 photometric observations by the end of the nominal (5 year) mission. We will refer to the traversal of a source across the \gaia focal plane as a transit.

The \gaia data are processed in several data centers across Europe, where each Coordination Unit (CU) performs specific data reduction and validation tasks \citep{Prusti2016}. Following this reduction and validation, there are a number of flags and consistency checks that are added as the data flows through the data reduction pipeline. 

\subsection{Further data processing}
Before we describe the statistical analysis applied in our search for fast transient events, we first briefly describe the flags and filtering steps that we will use to obtain a {\it cleaned} data sample of per CCD photometric measurements. Most of the filtering applied here concerns consistency checks for each transit; if there are flags indicating that the transit may be compromised, we remove it from the data associated to a source. Each source detected by \gaia has an ID, and likewise each transit also has an ID. For a more detailed description of the meaning of all the flags discussed below we refer to the \gaia DR1 documentation\footnote{https://gaia.esac.esa.int/documentation/GDR1/pdf/GaiaDR1\\$\_$documentation$\_$1.0.pdf} and Hodgkin et al. (in prep.). We note that we do not use GDR1 data as it does not contain the per CCD measurements. Instead, we use the database dedicated to the \gaia Science Alerts project to perform our analysis. A similar filtering procedure is also implemented in the \gaia Science Alerts pipeline (Hodgkin et al., in prep.). 

During the initial data treatment (IDT; \citealt{Fabricius2016}), flags that indicate whether the processing of the individual measurements have been performed nominally are provided. Transits can be blacklisted as bright star artefacts. When no spacecraft attitude is available to compute the sky positions or when either the astrometry or photometry are thought to be compromised because of irregular mission events, transits are also flagged. For example, in the early life of the \gaia mission, there was a decrease in the system throughput of the optical train due to contamination by water ice \citep{Evans2017}. All transits occurring while the spacecraft was affected are removed from the data.
Additionally, for our analysis we are conservative and discard transits that have more than 1 missing datapoint, as this implies that something has likely gone awry during the data processing chain (but did not necessarily get flagged). We check that the gate and window class configurations are compatible with each other and with the brightness of the object at AF1. Transits with missing windows, charge injections or near the CCD edges are all removed. Each time \gaia scans the same part of the sky, a cross match is performed to identify to which historic source ID (if any) the new observation belongs to. Flags are provided to keep track of ambiguous detections (i.e. sources that have more than one match candidate); we only keep transits that are clean matches, meaning that only one candidate was found during cross matching. This ensures that the source is isolated and there is no possible source confusion. 

\begin{figure*} 
\includegraphics[width=\textwidth, keepaspectratio]{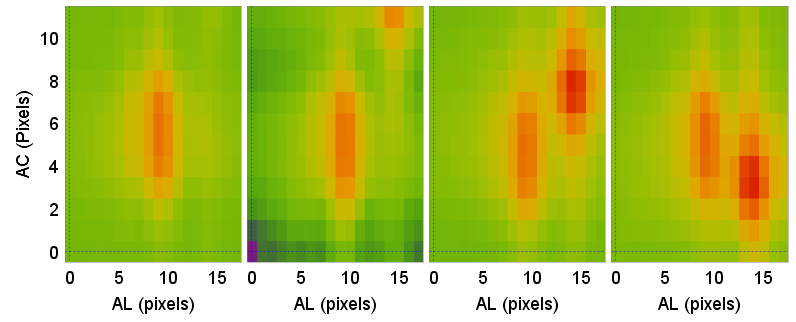}
    \caption{Example of a parasitic source transit. Each panel depicts a CCD row. The parasitic source comes in from above and passes within the (2D) window of another source, causing an artificial increase in source brightness. In this case the brightening lasts for 3 CCD crossings at the end of the transit. Because the object is relatively bright, a 2D window is available and the effect can be visually confirmed. However, the majority of faint sources have 1D windows and in that case we rely only on the g.o.f. values.}
 \label{fig:parasitic}
\end{figure*}

Owing to the spacecraft precession, sources in the \gaia FoV do not move perfectly parallel in the AL direction, but also move in the across-scan (AC) direction (perpendicular to the AL direction; this motion is called AC drift). An allocated window can move as much as 4.5 pixels in the AC direction during a single CCD transit \citep{Fabricius2016}. Because most windows are binned in the AC direction (for faint sources), and GDR1 only contains average flux measurements from a FoV transit, this does not affect the GDR1 measurements significantly \citep{Fabricius2016}. The amount of AC drift depends on the AC position of the source and the scanning angle, and is not the same for sources at the same location in the focal plane but in different FoVs. This shift changes during a revolution, and is updated regularly as a source crosses the focal plane. However, the two FoVs are affected differently, which can lead to non-rectangular windows and associated flux losses, or windows may suddenly enter in conflict with windows from the other FoV. If a window of a source in 1 FoV crosses over a window from a source in the other FoV, this can create apparent brightness variations at the CCD level dubbed {\it parasitic sources}. Because the effect depends on the relative positions of the windows, which are unlikely to be exactly the same for different transits (due to the specific scanning law \gaia employs and because windows are assigned without memory, or in other words the window position for a given source can be different for different transits), this can cause an apparent transient brightening in the lightcurve. We show an example of such an event in Figure \ref{fig:parasitic}.
If these events are short enough they can be rejected at the detection stage (Section \ref{sec:algorithm}). However, this effect can also create a monotonic increase in the source flux over the transit, which fits the criteria of our search algorithm. Such events are spurious but otherwise indistinguishable from real astrophysical transients based on the photometry alone. We identify such sources based on the so-called goodness of fit (g.o.f.), which is an image parameter diagnostic based on a $\chi^2$ fit of the line spread function\footnote{The line spread function is the 1-dimensional equivalent of the traditional point spread function. Faint sources are allocated 1-dimensional windows by \gaia to reduce the data volume.} (LSF). When two windows overlap, the sum of the two individual LSFs does not correspond to the LSF of a single source (because the sum of LSFs changes over time). As a consequence, overlapping windows cause a deterioration of the g.o.f. over time, and can be identified as transits that have anomalous or highly variable g.o.f. values. These transits are suspect and we remove them from the data.

As a final step, we require a source to have at least 5 clean transits. This is necessary because our outlier detection algorithm depends on the statistical properties of the data; too few datapoints will make the analysis potentially unreliable. It also implies that our search is currently only sensitive to sources with a historic lightcurve, and not to transient sources that appear new to the \gaia source catalogue. The filtering as described above yields a data set that is free of known on-board and data processing artefacts, and forms the basis of the rest of our analysis. 
For a clear overview, we summarize the filtering steps applied for each transit as a bulleted list:
\begin{itemize}
\item Not blacklisted by IDT
\item Less than 1 missing datapoint
\item Compatible gate, window class and source brightness
\item No missing windows
\item Not near charge injection
\item Not near CCD edge
\item Clean cross-match
\item Well-behaved g.o.f.
\end{itemize}

In addition, there are a number of situations that can introduce apparent brightness variations in the per CCD \gaia data. These can be identified by investigating the environment of a source as they are generally related to close neighbouring sources or bright stars. 
\subsection{Environment analysis}
Any source that is not the dominant source (defined here as the brightest source by 1 mag or more) within a 2 arcsec radius is rejected as it may be confused with its neighbour, leading to an artificial flux variation due to a bad cross match. Similarly, if stars brighter than $G$\,=\,12 mag are within 10 arcsec of a source, it is rejected due to the large probability of contamination. We also reject sources for which the position of the brightening transit is clearly inconsistent with the scatter on the quiescent source position, indicating that the variability is potentially caused by another source nearby. 

In addition, diffraction spikes around relatively bright stars are known to cause problems in the lightcurves of \gaia objects, either because of spurious source detections along the diffraction spikes or by artificially increasing the source flux if a diffraction spike overlaps with a source window. 
For each candidate transient we search a circular region with radius 1 arcmin centered on the source for stars brighter than $G$\,=\,12 mag. If such a bright source is present, we use the scan angle of the transit to estimate whether a diffraction spike could be the cause of the increased brightness (diffraction spikes occur in preferential directions; see Hodgkin et al. in prep. for a more detailed description). We reject candidate fast transients that are potentially located on a diffraction spike. 

\subsection{Contamination from the secondary FoV}
The analysis described above aims to exclude all known data artefacts as well as potential environmental issues such as contamination due to nearby neighbouring stars or parasitic sources. Due to the organistation of the \gaia data, this is fairly trivial and computationally not very expensive to perform for the same FoV. However, because both FoVs pass over the same detector plane but are processed independently, it is possible that environmental issues arise due to sources on the alternate FoV (i.e. the FoV where the source is {\it not} located), for example the presence of a bright star and associated diffraction spikes on the alternate FoV. To perform this analysis, it is necessary to reconstruct the window positions and velocities of each source in the alternate FoV; this information is currently neither readily available nor easy to reconstruct given that the accuracy of the source position reconstruction must be at the $\sim$\,pixel level. This is beyond the scope of the current work, but will be performed in the future.
To mitigate this effect to first order, we determine the Galactic coordinates of the center of the opposing FoV for the transit that triggers our detector, and reject transients for which this corresponds to locations near the Galactic plane (at Galactic latitude $|$b$|$\,$\leq$\,15$^{\circ}$) as the source density is expected to be high, hence the probability of such contamination occurring increases.

\section{Outlier detection}
\label{sec:algorithm}
\begin{figure*} 
\begin{minipage}{0.495\textwidth}
\includegraphics[width=\textwidth, keepaspectratio]{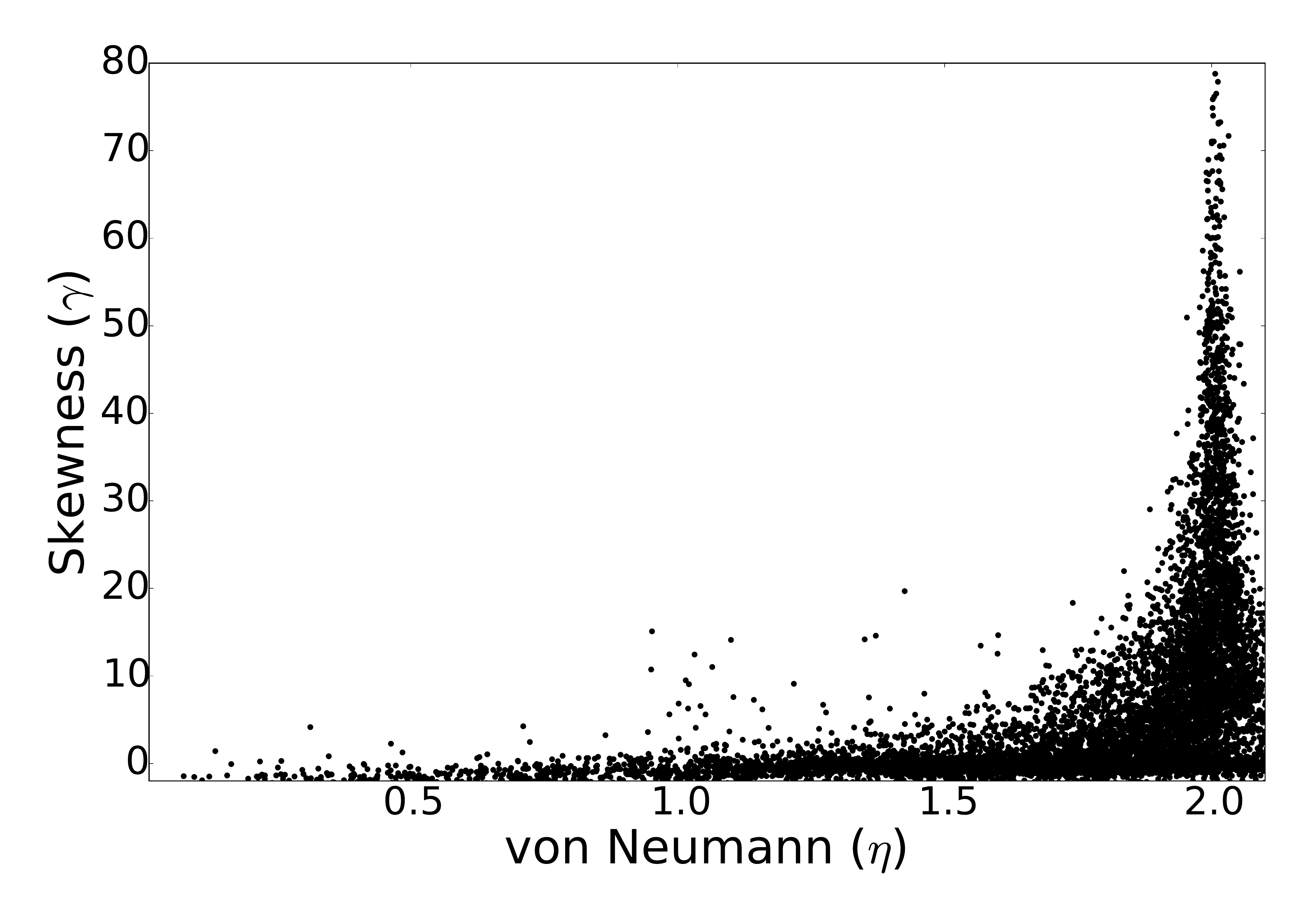}
\end{minipage}
\begin{minipage}{0.495\textwidth}
\includegraphics[width=\textwidth, keepaspectratio]{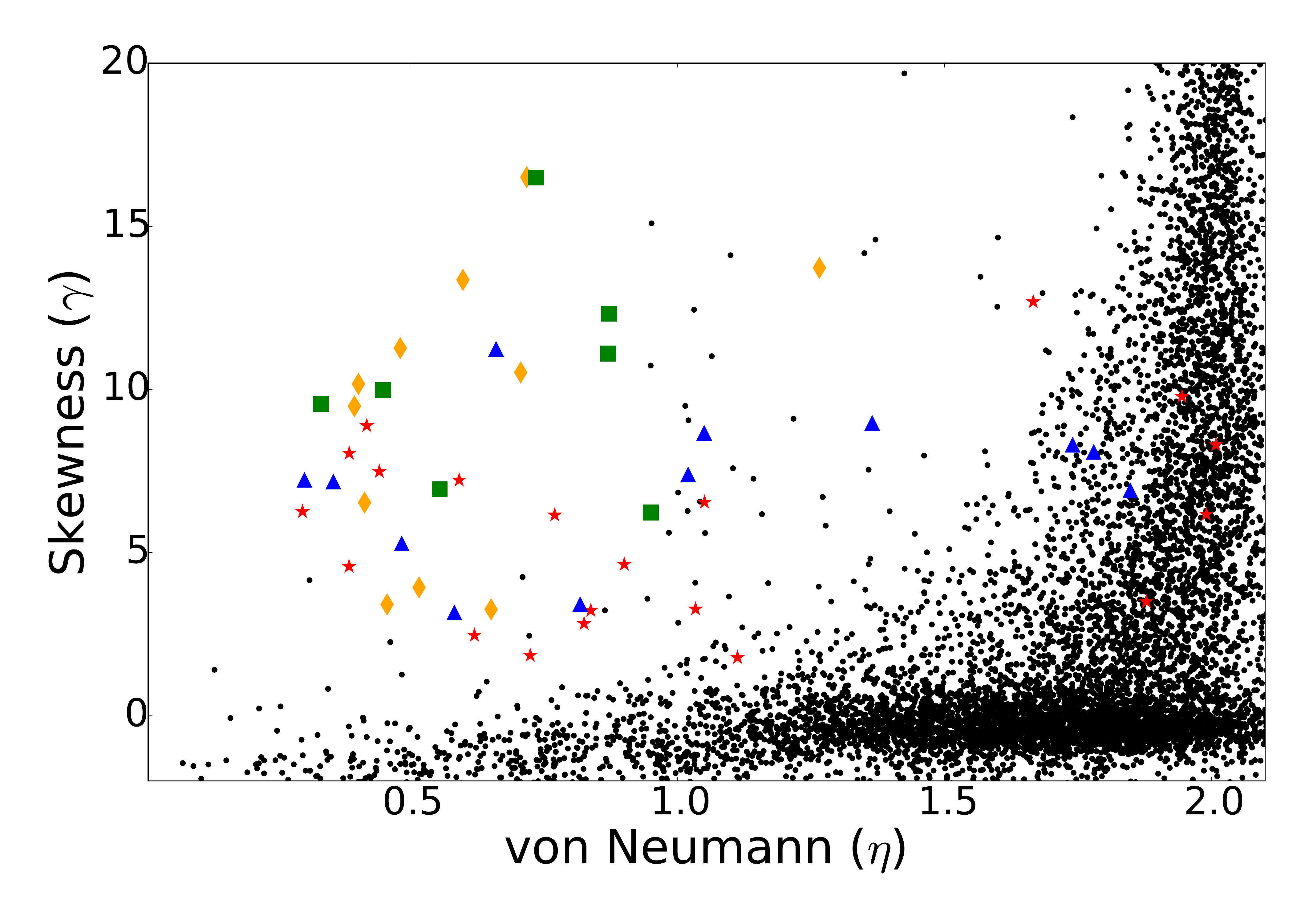}
\end{minipage}
    \caption{Left: von Neumann and skewness diagram of HEALpixel 2893, containing $\sim$\,9600 sources. Constant or periodically variable lightcurves are located at the locus around $\gamma$\,=\,0, while lightcurves containing no or non-sequential outliers tend towards $\eta$\,=\,2. The outliers from these two loci represent the lightcurves containing potential transients. Right: Illustration of the positions of the artificial transients in ($\gamma$, $\eta$) space. Different markers indicate the amplitude of the transients, ranging from $\leq$\,1.5 mag (red stars), between 1.5 and 2 mag (blue triangles), between 2 and 2.5 mag (orange diamonds) up to $\geq$\,2.5 mag (green squares). }
 \label{fig:gamma_eta}
\end{figure*} 
The aim of our work is to robustly identify transient events on short timescales ($\leq$\,2 hours). Because the method needs to be scalable to the large data volume that \gaia collects every day, we choose to use the standardized skewness ($\gamma$) and the von Neumann statistic ($\eta$) of the lightcurve (see e.g. \citealt{Pricewhelan2014, Wyrzykowski2016} for applications to microlensing searches). The former metric allows us to define a selection criterion to remove sources exhibiting stochastic / periodic variability, while a selection criterion based on $\eta$ allows us to select smooth variations (in contrast with single data point outliers induced by e.g. cosmic rays). 
The von Neumann statistic is defined as the ratio of the mean square successive difference to the variance \citep{Neumann1941} 
\begin{equation}
\eta = \frac{\delta^2}{s^2} = \frac{\frac{1}{n-1} \sum\limits_{j=1}^{n-1}{(x_{j+1} - x_{j})^2}}{s^2}
\end{equation}
A strong positive serial correlation (trend) between datapoints will lead to a low von Neumann statistic, and indicates smooth variability as opposed to single (or no) outliers, which will lead to $\eta$\,$\sim$\,2 values. Similarly, near constant or periodically variable sources will have a low skewness, while the presence of outliers increases the skewness of a lightcurve. We use the combination of these two metrics to characterize each lightcurve, and the problem of finding transient events is reduced to efficiently determinining the parameter space which contains lightcurves exhibiting transient events. The analysis is performed at HEALpixel (HP) level 5 \citep{Gorski2005}, which divides the sky in 12288 pixels with an approximate size of 3.36 square degrees per pixel. We apply the analysis to HPs (nested ordering) with following IDs: 285, 859, 2893, 2904, 5731, 7788 and 8286. These HPs were selected to cover relatively uncrowded regions (between 9600 and 83000 clean sources per HP), to avoid cross match problems associated to regions with high stellar densities. They span an area of approximately 23.5 square degrees on the sky, and the typical (uncleaned) lightcurve contains about 40\,--\,50 transits observed during the nominal mission. The exception is HP859, which has been observed roughly 90 times by \gaia during the nominal scanning law mission.
We show an example of a ($\gamma$, $\eta$)-diagram in Figure \ref{fig:gamma_eta} for HP2893. It is clear from the figure that the scatter in the skewness increases as $\eta$ increases. These lightcurves most likely have only a few outlying (potentially non-consecutive) datapoints given their high von Neumann values and therefore we aim to reject them as candidate fast transients due to the risk of contamination from spurious events.

\subsection{Artificial transient events}
The precise shape of the ($\gamma$, $\eta$) diagram is different for each HP. In order to gain some insight into the relevant parameter space, we add short, artificial transient events to existing \gaia lightcurves. To this end, we randomly select 50 lightcurves and inject a rising transient with the following properties:
\begin{itemize}
\item The starting time of the transient is random
\item The transient duration is between 4 and 10 datapoints (18\,--\,45 s)
\item The peak amplitude is 0.3 to 3.5 mag above the quiescent brightness
\item The transient is back in quiescence in the next \gaia observation
  \end{itemize} 
We show the resulting lightcurves with respect to the clean \gaia lightcurves in the right panel of Figure \ref{fig:gamma_eta}. The figure shows that part of the lightcurves with artificial transients occupy a distinct parameter space in the diagram. The transient lightcurves generally occupy the parameter space bounded by (roughly) $\eta$\,$\leq$\,0.75. This is a consequence of the minimum duration of the transient events that we injected in the lightcurves. Because our goal is to assess the potential of the per CCD \gaia data to identify fast transient events, we choose to minimize the false positive identifications by requiring $\eta$\,$\leq$\,0.75 for the rest of our analysis. 

While we use $\eta$ as a rejection criterion for likely spurious, short duration events, the skewness of the lightcurve is mainly influenced by the amplitude of the variability (although there is also some dependence on event duration). In Figure \ref{fig:gamma_eta} (right panel) we illustrate how the amplitude of the transients influences their position in the diagram. Black dots represent the cleaned \gaia data, while red stars, blue triangles, orange diamonds and green squares represent transients with increasing amplitudes ranging from $\leq$\,1.5 mag to $\geq$\,2.5 mag in steps of 0.5 mag. 

\begin{figure} 
\includegraphics[width=0.5\textwidth, keepaspectratio]{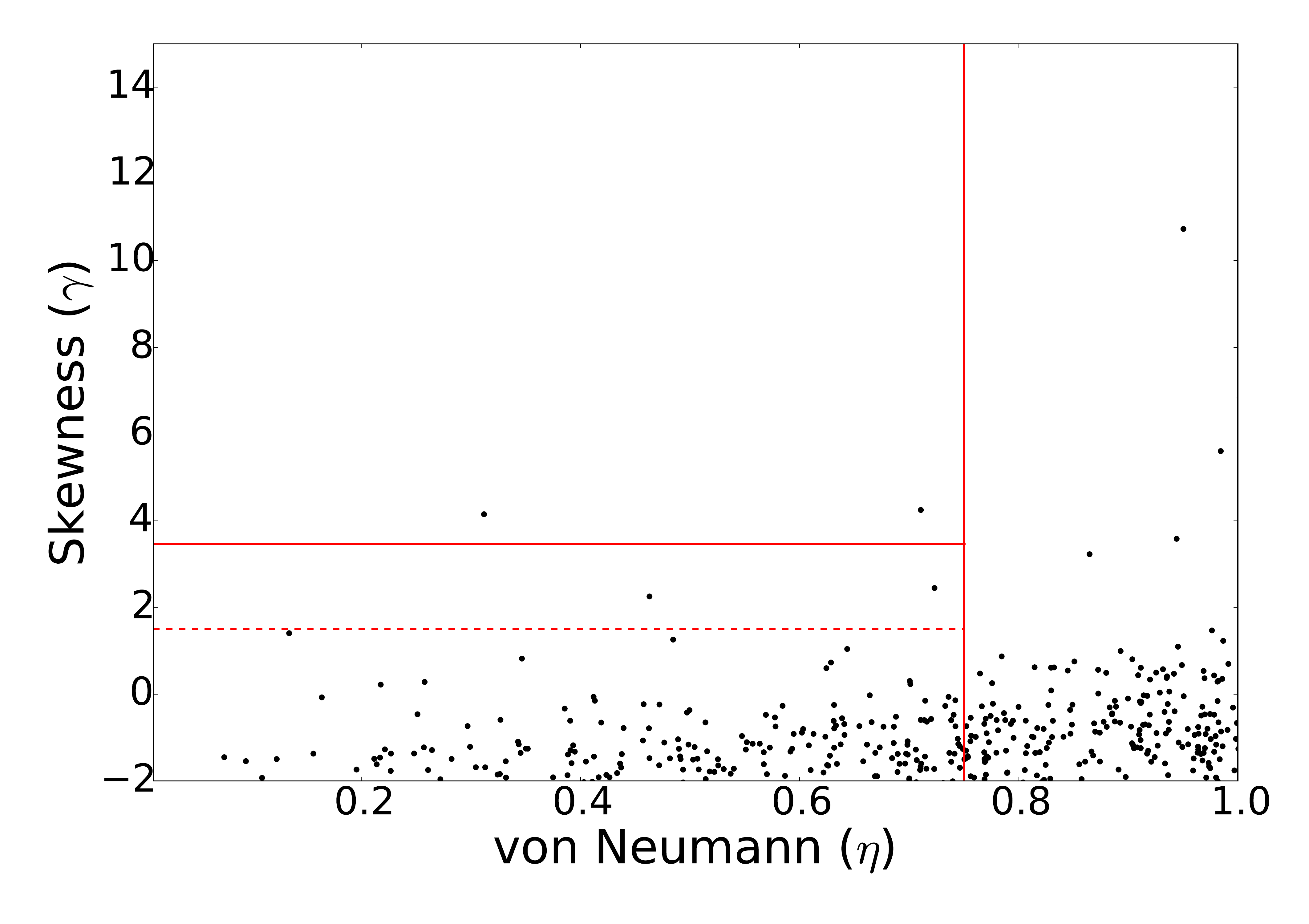}
    \caption{Cleaned \gaia data of HP2893. The red horizontal lines indicate selection criteria of 3\,$\sigma$ (dashed) and 5\,$\sigma$ (solid); the vertical red line indicates a cut at $\eta$\,$\leq$\,0.75. The two sources in the upper left quadrant (defined by the solid lines) are identified as candidate fast transients.}
 \label{fig:outliersdiagram}
\end{figure}

We exploit the isolation of the lightcurves in the skewness metric to robustly select outliers. We use the mean and standard deviation of the skewness distribution of the cleaned \gaia lightcurves with $\eta$\,$\leq$\,0.75 to set the selection threshold. In Figure \ref{fig:outliersdiagram} we illustrate the selection boundaries for respectively 3\,$\sigma$ and 5\,$\sigma$ using the dashed and solid horizontal lines. We retain all sources that are $\geq$\,5\,$\sigma$ from the mean as robust outliers, and will investigate them in more detail.

\begin{figure} 
\includegraphics[width=0.5\textwidth, keepaspectratio]{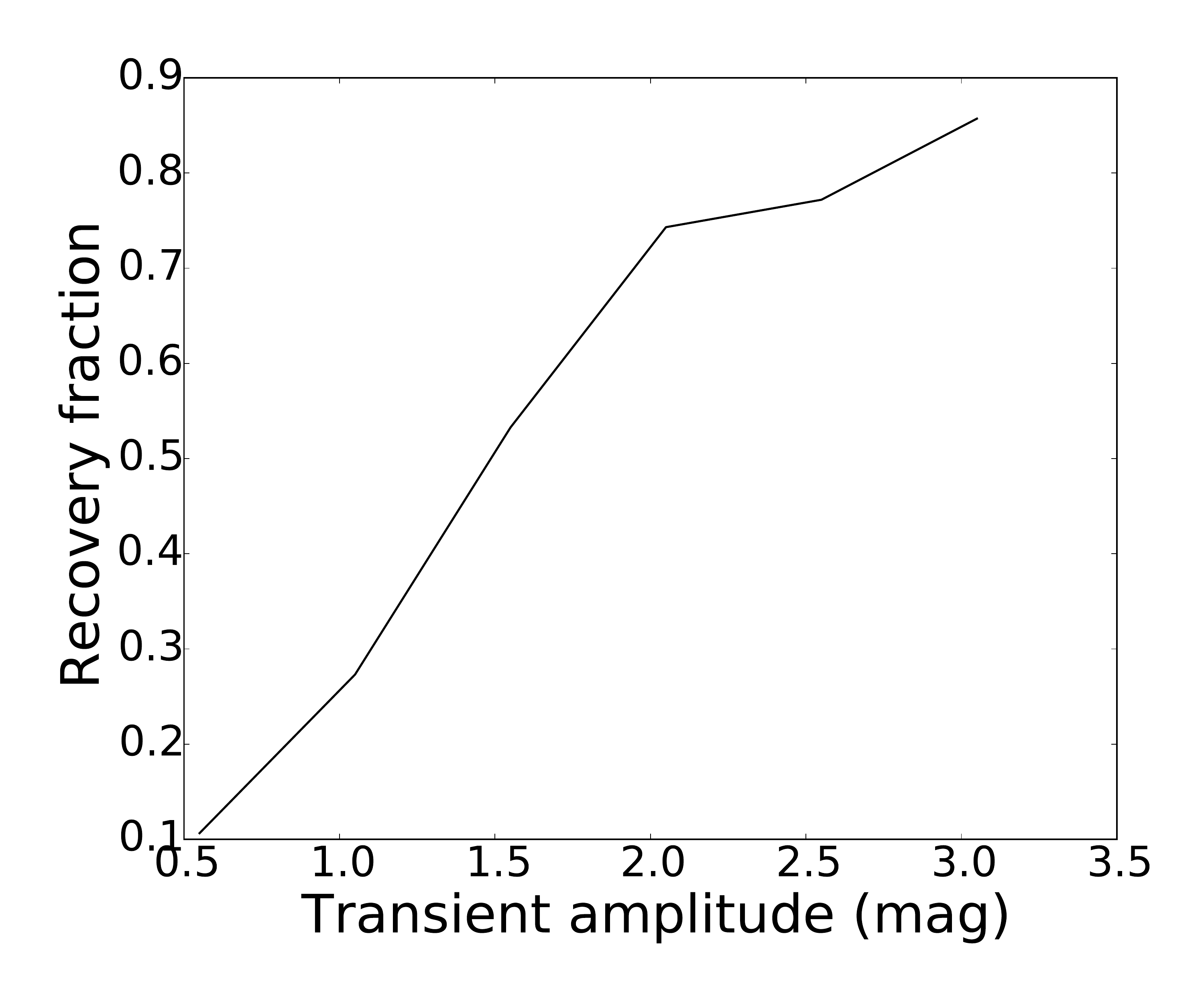}
    \caption{Recovery fraction of artificial transients as a function of transient amplitude for HP2893.}
 \label{fig:recoveryfraction}
\end{figure}

To obtain an estimate of the completeness of the method, we inject transients in 1000 lightcurves of a given HP. The resulting recovery fraction, based on the thresholds detailed above, is shown in Figure \ref{fig:recoveryfraction} as a function of the transient amplitude with respect to the baseline for HP2893. We find that, as expected, we recover an increasing fraction of outliers with increasing transient amplitude. The low recovery fraction at low peak amplitudes is due to the conservative constraints applied in the analysis to minimize the number of false positives at the expense of a lower detection probability. In future work we will aim to improve the recovery fraction of low amplitude transients (but see Section \ref{sec:fasttransientresults} for an example of such an event).

\section{Results}
\label{sec:fasttransientresults}
In the 7 HP trial dataset, eight candidate fast transients survive our filtering and selection procedures, and we list their coordinates and \gaia magnitudes together with the duration and amplitude of the transients in Table \ref{tab:candidatefasttransients}. We cross match these sources with the SDSS and Pan--STARRS catalogues when available, and provide the $r^{\prime}$\,--\,$i^{\prime}$ colour index. This index can be used to estimate the spectral type of late type stars \citep{Hawley2002}. When the source position is located within the survey footprints, we visually inspect the close environment of the candidates to confirm that an astrophysical source is present in these surveys.
\begin{figure*} 
\begin{minipage}{0.495\textwidth}
\includegraphics[width=0.84\textwidth, keepaspectratio]{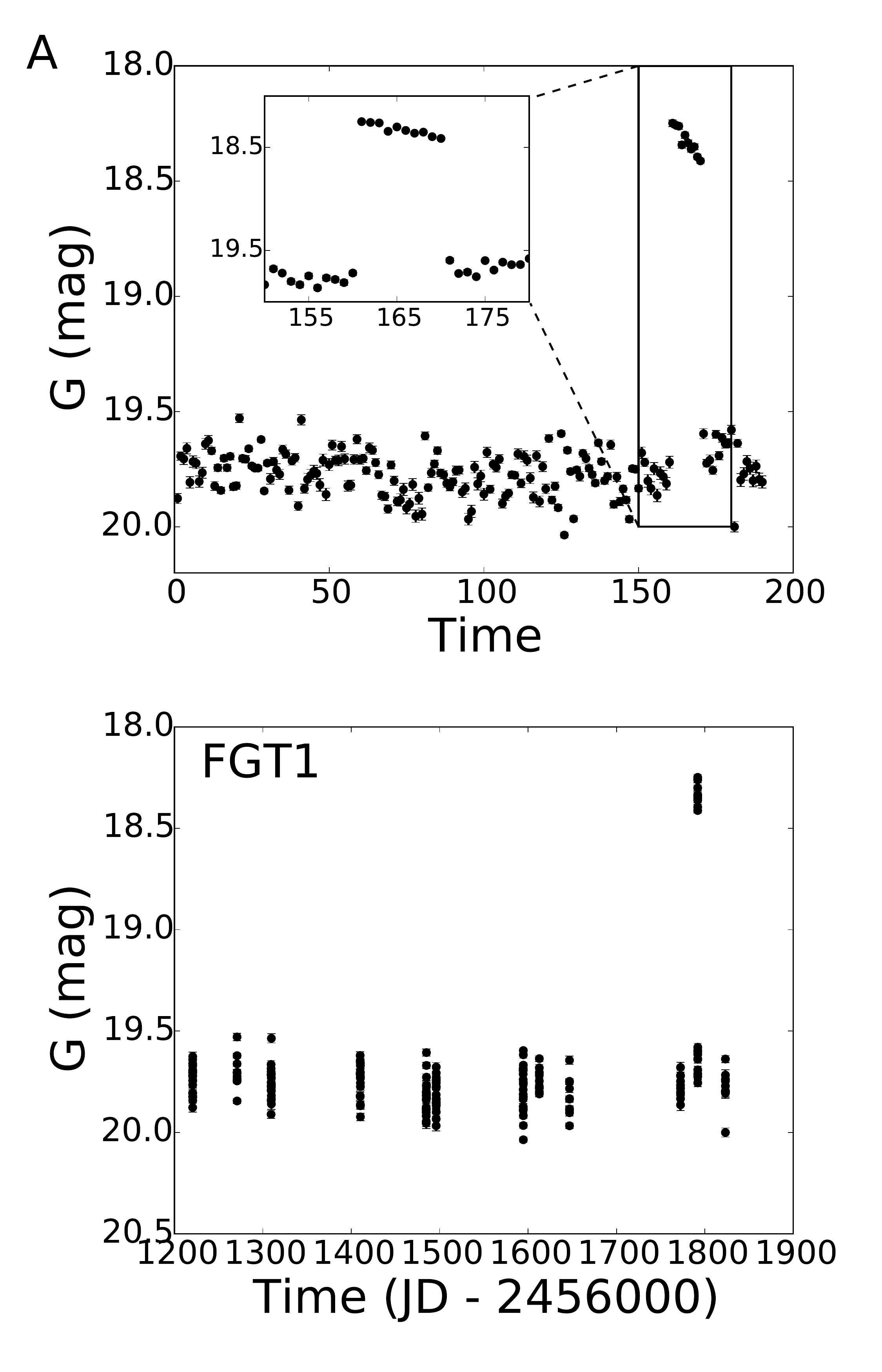}
\includegraphics[width=0.84\textwidth, keepaspectratio]{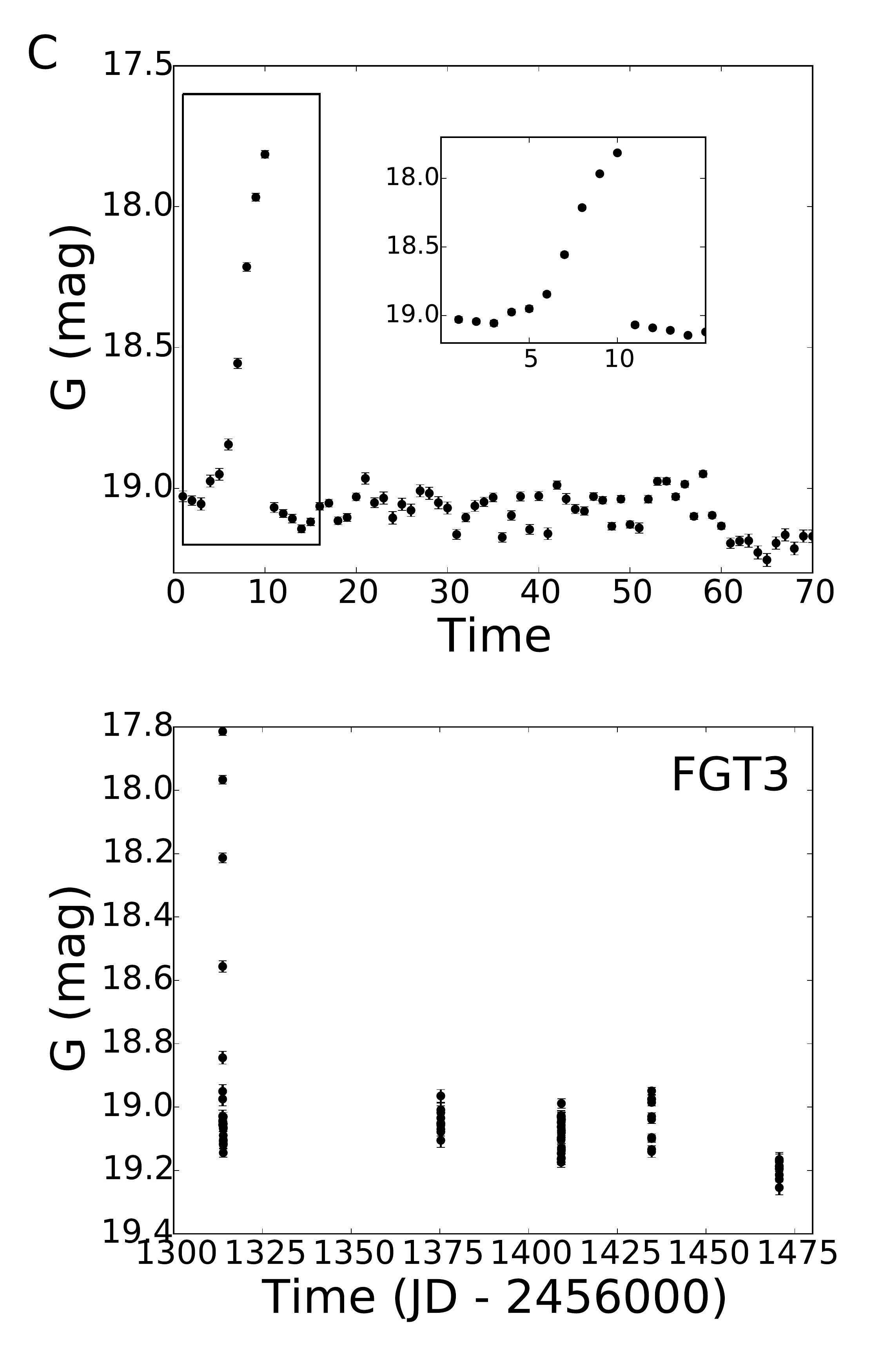}
\end{minipage}
\begin{minipage}{0.495\textwidth}
\includegraphics[width=0.84\textwidth, keepaspectratio]{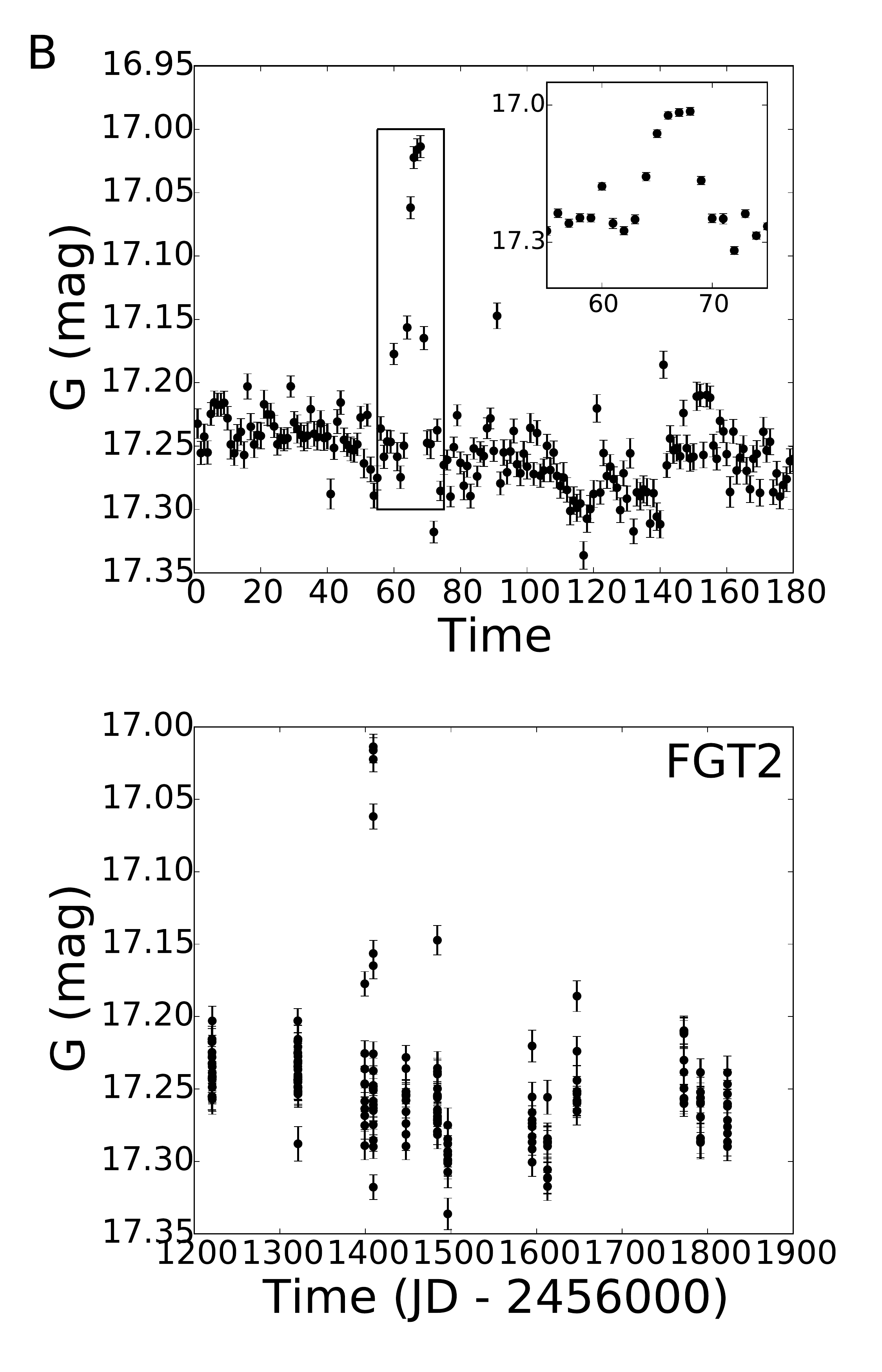}
\includegraphics[width=0.84\textwidth, keepaspectratio]{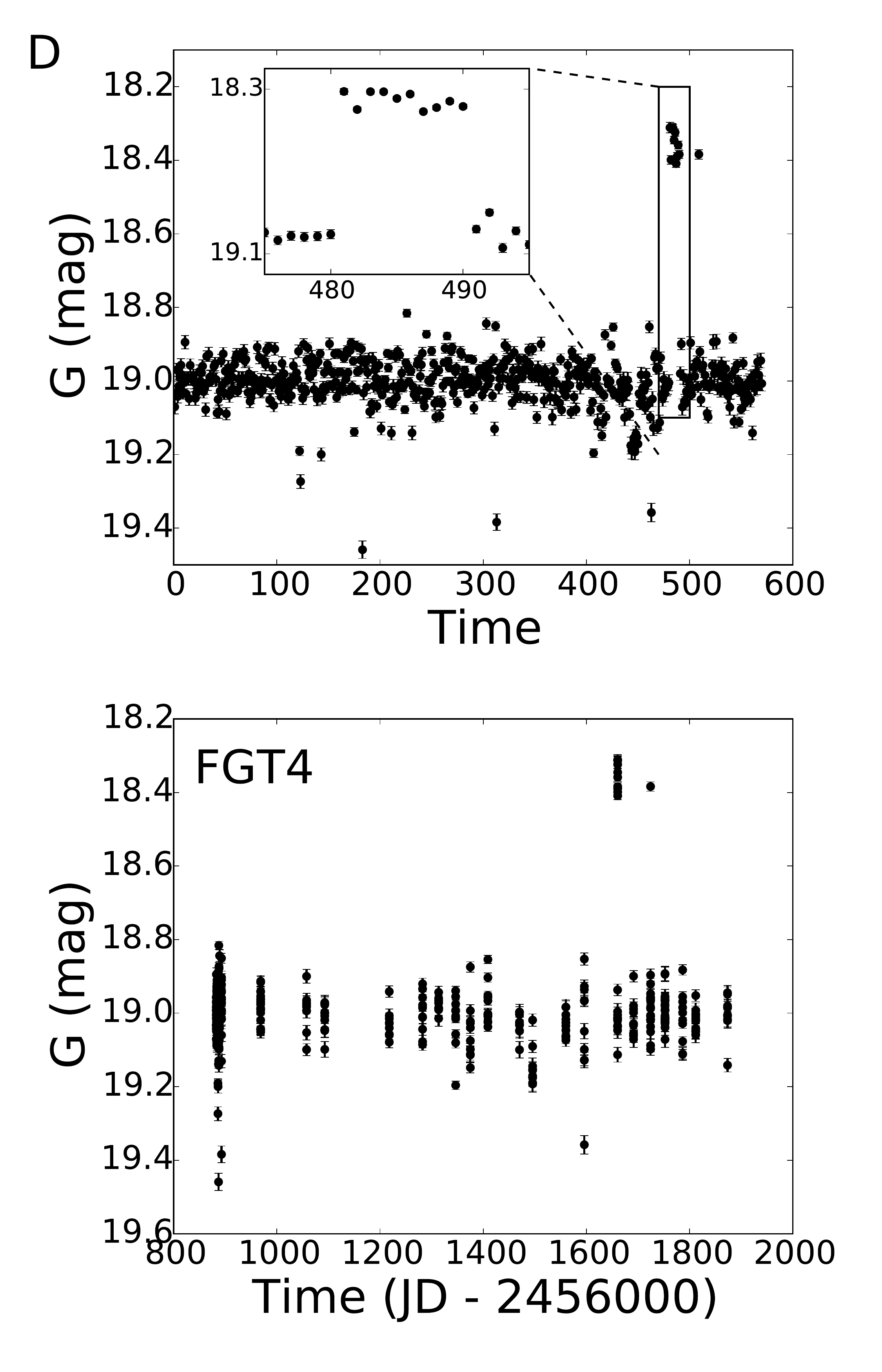}
\end{minipage}
    \caption{Lightcurves of FGT1 (panel A) through FGT4 (panel D). Each panel consists of 2 figures: the top figure shows the concatenated lightcurve for display purposes, with a running number per CCD crossing (i.e. time increases with running number). 10 measurements constitute 1 transit. The inset shows a zoom-in to the transient event. The bottom panel shows the actual time spacing of the observations, in Julian days. The data of the outlying transits is presented in tabular form in the Appendix.}
 \label{fig:cft}
\end{figure*} 
\begin{table*}
 \centering
  \caption{Properties of the candidate fast \gaia transients (FGTs). The events above the line are most likely real astrophysical events, while for the events below the line there is the possibility that the brightness increase is due to (unknown) SSO crossings of the window assigned to the source. Coordinates are given in decimal degrees. RA$_{\rm lit}$ and Decl$_{\rm lit}$ are the position of the optical counterpart (if available). $G$ is the mean \gaia magnitude, with the number in brackets giving the uncertainty on the last digit (determined as the standard deviation of the historic lightcurve). $\Delta$mag is the amplitude of the transient with respect to the mean. The duration for sources marked with an asterisk are lower limits. $r^{\prime}$\,--\,$i^{\prime}$ is the SDSS (S) or Pan--STARRS (PS) colour index. Comments are tentative classifications where possible, and indicate whether a \gaia sightline was near (within 15$^{\circ}$) the ecliptic plane (E).}
  \begin{tabular}{cccccccccc}
  \hline
  Name & RA$_{\gaia}$ ($^{\circ}$) & Decl$_{\gaia}$ ($^{\circ}$) & RA$_{\rm lit}$ ($^{\circ}$)& Decl$_{\rm lit}$ ($^{\circ}$)& $G$ (mag) & $\Delta$mag & Duration (sec) & $r^{\prime}$\,--\,$i^{\prime}$ & Comments \\
  \hline
FGT1 & 72.38784& --0.79425 &72.38784 & --0.79420 & 19.7(3) & 1.4 & 45$^{\star}$ & 2.1 & M dwarf, S\\
FGT2 & 253.32474 &1.99771&253.32475  & 1.99773& 17.25(3) & 0.25 & 36 & 0.21 & S \\
FGT3 & 251.23489 &64.47513& 251.23514& 64.47514& 19.0(1) & 1.25 & 32$^{\star}$ & 1.55 & M dwarf, S, E\\ 
FGT4 &73.93111& --64.71406 & -- & -- & 19.0(1) & 0.7 & 45$^{\star}$& -- & K/M giant, E\\\hline
FGT5 & 74.96236 & 33.26740 & 74.96237 & 33.26740 & 19.1(1) &0.7 & 27$^{\star}$ & 0.69 & PS, E \\
FGT6 & 75.21896 & 33.36786 & 75.21896 & 33.36784 & 18.55(8) & 0.6 & 27& 0.59 & PS, E\\
FGT7 & 74.93751& 31.87202& 74.93753 & 31.87201 & 19.52(7) & 1.5 & 14 & 0.45 & PS, E \\
FGT8 & 71.45835& --62.85321& --& --& 18.86(4) & 0.35 & 14 & -- & E \\
\hline
    \end{tabular}
  \label{tab:candidatefasttransients}
\end{table*}

For 6 out of 8 sources (the exceptions being Fast \gaia Transient (FGT) 1 and FGT2), one of the \gaia FoVs is pointing near the Ecliptic plane at the time of the transient event, implying that some of these brightness variations may be caused by solar system objects (SSOs) passing through the FoV. We will argue below that for at least two of these sources (FGT3 and FGT4), this is unlikely to be the case and these are robust astrophyical transients. We use the Skybot service \citep{Berthier2006} to check if any known SSOs could be the cause of the brightness variations. In two cases (FGT5 and FGT7) there is a known asteroid within 500 arcsec, however, the SSO optical brightnesses ($V$\,=\,20.7 and 21.8 mag, respectively) cannot explain the increased brightness of 0.7 and 1.5 mag respectively. Therefore we conclude that no known SSO is the cause of the transient events discussed here. We now briefly discuss the four most robust candidates in some more detail. We dub these robust because there is either archival data available that corroborates with the interpretation of the event, or they are not located near the Ecliptic plane and it is therefore highly unlikely that the transient is due to an unknown SSO.

We show the lightcurve of FGT1 in Figure \ref{fig:cft} (panel A). This source is coincident with a red star in the SDSS catalogue. Its $r^{\prime}$\,--\,$i^{\prime}$ colour\,=\,2.1 indicates a spectral type of M5--6V \citep{Hawley2002}. The brightness is monotonically declining during the outlying transit at a rate of $\sim$\,0.2 mag min$^{-1}$. We therefore tentatively classify this transient as a flare on a M-dwarf star. 

FGT2 coincides with a star with colour index $r^{\prime}$\,--\,$i^{\prime}$\,=\,0.21, which does not point towards a late type star. The lightcurve (Figure \ref{fig:cft} panel B) shows a transient with a duration of 8 CCD crossings or 36 seconds, and the lightcurve is slightly asymmetric with the decline faster than the rise time. The peak amplitude is $\sim$\,0.25 mag. This transient illustrates the sensitivity to transient events on timescales below 40 seconds, which is a unique capability of {\it Gaia}.

FGT3 can be cross matched in the SDSS catalogue with an isolated red stellar object with $r^{\prime}$\,--\,$i^{\prime}$\,=\,1.55, indicative of a M4V spectral type \citep{Hawley2002}. The lightcurve (Figure \ref{fig:cft} panel C) starts rising within a transit, and reaches an amplitude of 1.25 mag in 32 seconds. Although the secondary FoV for the transit containing the transient is near the Ecliptic plane (at Ecliptic latitude of --12 degrees), the coincidence with an M type dwarf star leads us to classify it as a robust transient event which is likely a flaring M-dwarf. The source is back in quiescence during the following \gaia scan 106 minutes later. The lightcurve illustrates that \gaia can provide constraints for the rate of rise of transient events. Such constraints could be used to obtain a preliminary classification, and in addition may provide insights into the physics underlying the brightness variations.

FGT4 does not overlap with either SDSS or Pan-STARRS but is located along the line of sight towards the Magellanic Clouds (MCs). It is detected in the 2MASS \citep{Skrutskie2006} and WISE \citep{wright2010} catalogues with J\,=16.67\,$\pm$\,0.06, H\,=\,15.97\,$\pm$\,0.08, K$_{\rm S}$\,=\,16.1\,$\pm$\,0.15, W1\,=\,15.95\,$\pm$\,0.05 and W2\,=\,15.71\,$\pm$\,0.09 mag. The colours are consistent with a K/M type giant star; assuming it is located in the MCs its absolute K-band magnitude is $\sim$\,--2.5, consistent with this interpretation. The lightcurve is shown in Figure \ref{fig:cft} (panel D), and shows an entire transit outlying from the quiescent source brightness. Although for this transient, the secondary FoV is located near the Ecliptic plane (Ecliptic latitude of 11 deg), we deem it unlikely that a moving object would contaminate a complete transit. This event could be a flare on a late type giant star.

For the remaining four sources, one of the FoVs is close to the Ecliptic plane. These sources cannot be clearly identified with a late type star. Although this is clearly not a necessary requirement, to confirm the astrophysical nature of these events requires multi wavelength or time resolved follow up observations.

Our results illustrate that the method is sensitive to both short (tens of seconds) and longer (minutes to hours) timescale brightness variations. Moreover, our method allows us to select sources that exhibit variations of the order of 0.3 mag. This reflects and confirms the high quality of the individual \gaia photometric CCD measurements.

We can roughly estimate the expected rate of fast transients discovered by \gaia by assuming that our 7 HP trial dataset is representative of the rest of the sky. In that case, we take into account to first order the area on the sky that is lost by excluding 15 degrees below and above the Galactic plane. This amounts to roughly 10000 square degrees, leaving 31000 square degrees of the sky that can be searched. In our work we find 4\,--\,8 transient candidates in 23.5 square degrees, and taking into account that \gaia scans $\sim$\,1200 square deg per day we arrive at a rate of 4\,--\,9 fast \gaia transients per day. This order of magnitude estimate does not take into account the area on the sky that is eliminated around bright stars and is likely an overestimate.

\subsection{Future improvements}
The implementation of a random forest classifier, based on the low resolution spectra that are measured quasi-simultaneously with the photometry, can help to classify the newly obtained transients into broad classes \citep{Blagorodnova2014}. \gaia has the unique feature that each transit has an associated Bp and Rp spectrum, taken during or very close to the transient event. This is especially useful for short duration transient events, as they are likely to be back in quiescence by the time of dissemination (typically the data downlinking and processing takes 48 hours). These spectra may thus contain vital spectral information in addition to the colour information they provide regarding the nature of the transient events. The calibration of these spectra is envisaged to be performed in the near future, and may help to both reject spurious events as well as aid the (preliminary) classification of newly discovered fast transients.

In addition, improvements in the astrometry will be implemented in the \gaia Science Alerts database in the near future. Currently, the source positions are based on a first pass of the On-ground Attitude determination (OGA1) during IDT. In addition, a second iteration is performed during the later stages of data processing, leading to improved source positions (OGA2). The accuracy of the OGA2 source positions is 1\,--\,2 orders of magnitude better than OGA1. 
Although for a newly identified transient only the OGA1 position will be available shortly after detection, the implementation of the OGA2 positions for the historic transits will help to reject brightness variations that are inconsistent with the quiescent source position and thus improve the rejection of spurious events.

Finally, the reconstruction of the sky positions of sources on the secondary FoV will increase the confidence in detections of sources similar to FGTs 5 through 8, where it is currently unclear whether they are real or could be due to objects in the second FoV. Using the reconstructed second FoV will allow us to reject or confirm the astrophysical nature of these transient events and therefore will aid in the automatisation of the detection process.

\section{Summary}
\label{sec:summaryandfuture}
In this work, we have explored the feasibility of using the \gaia CCD data to search for short timescale transient events. \gaia provides sparsely sampled, high time resolution ($\sim$\,4.5 s) lightcurves for all sources brighter than $G$\,=\,20.7 mag over the whole sky. Such timescales are virtually unexplored at optical wavelengths; \gaia provides a unique data set to explore this parameter space. We present the \gaia data and filtering steps to obtain a clean sample of sources, to which we apply a statistical method based on the skewness and von Neumann statistics (indicators for transient and positive correlations in the lightcurve) to select candidate fast transient events. Care is taken to account for all the known artefacts present in the data, as well as environmental issues that may lead to spurious results. The method is sensitive to identifying variability on timescales ranging from 10 s of seconds to hours. Moreover, the detection of relatively low amplitude (down to 0.3 mag) variability shows that the photometric precision of the CCD data is excellent. We detect 8 candidate fast transient events, of which we tentatively classify two as flares on M-dwarf stars and one as a flare on a giant star based on archival data.

The complete set of CCD photometric data from \gaia used in this work will be released to the public in GDR4\footnote{for the full data release plans, see https://www.cosmos.esa.int/web/gaia/release}.

The implementation of an automated pipeline for fast transient detection within the existing \gaia Science Alerts framework with minimal user intervention is feasible, and such a module can extend the timescales at which \gaia is sensitive to transients down to 10s of seconds. We conclude that \gaia will open a new window on the fast transient sky.
\section*{Acknowledgements}
TW acknowledges support from European Science Foundation Exchange Grant 4851 within the framework of Gaia Research for European Astronomy Training (GREAT-ESF). PGJ and ZKR acknowledge support from European Research Council Consolidator Grant 647208.
This work has made use of results from the European Space Agency (ESA) space mission Gaia, the data from which were processed by the Gaia Data Processing and Analysis Consortium (DPAC). Funding for the DPAC has been provided by national institutions, in particular the institutions participating in the Gaia Multilateral Agreement. 
The authors are current or past members of the ESA Gaia mission team and of the Gaia DPAC. This work has received financial
support from the European Space Agency in the framework of the Gaia project; the European Community’s Seventh Framework Programme (FP7-SPACE-2013-1)
under grant agreement no. 606740; the Netherlands Research School for Astronomy (NOVA) and the Netherlands Organisation for
Scientific Research (NWO) through grant NWO-M-614.061.414; the Swiss State Secretariat for Education, Research, and Innovation through the ESA PRODEX programme; the Mesures d’Accompagnement, and the Activit\'{e}s Nationales Compl\'{e}mentaires; the Swiss National Science Foundation; and the UK Space Agency, the UK Science and Technology Facilities Council.
\bibliographystyle{mnras.bst}
\bibliography{bibliography_fasttransients.bib}

\begin{thebibliography}{}
\makeatletter
\relax
\def\mn@urlcharsother{\let\do\@makeother \do\$\do\&\do\#\do\^\do\_\do\%\do\~}
\def\mn@doi{\begingroup\mn@urlcharsother \@ifnextchar [ {\mn@doi@}
  {\mn@doi@[]}}
\def\mn@doi@[#1]#2{\def\@tempa{#1}\ifx\@tempa\@empty \href
  {http://dx.doi.org/#2} {doi:#2}\else \href {http://dx.doi.org/#2} {#1}\fi
  \endgroup}
\def\mn@eprint#1#2{\mn@eprint@#1:#2::\@nil}
\def\mn@eprint@arXiv#1{\href {http://arxiv.org/abs/#1} {{\tt arXiv:#1}}}
\def\mn@eprint@dblp#1{\href {http://dblp.uni-trier.de/rec/bibtex/#1.xml}
  {dblp:#1}}
\def\mn@eprint@#1:#2:#3:#4\@nil{\def\@tempa {#1}\def\@tempb {#2}\def\@tempc
  {#3}\ifx \@tempc \@empty \let \@tempc \@tempb \let \@tempb \@tempa \fi \ifx
  \@tempb \@empty \def\@tempb {arXiv}\fi \@ifundefined
  {mn@eprint@\@tempb}{\@tempb:\@tempc}{\expandafter \expandafter \csname
  mn@eprint@\@tempb\endcsname \expandafter{\@tempc}}}

\bibitem[\protect\citeauthoryear{{Bauer} et~al.,}{{Bauer}
  et~al.}{2017}]{Bauer2017}
{Bauer} F.~E.,  et~al., 2017, \mn@doi [\mnras] {10.1093/mnras/stx417}, \href
  {http://adsabs.harvard.edu/abs/2017MNRAS.467.4841B} {467, 4841}

\bibitem[\protect\citeauthoryear{{Berger} et~al.,}{{Berger}
  et~al.}{2013}]{Berger2013}
{Berger} E.,  et~al., 2013, \mn@doi [\apj] {10.1088/0004-637X/779/1/18}, \href
  {http://adsabs.harvard.edu/abs/2013ApJ...779...18B} {779, 18}

\bibitem[\protect\citeauthoryear{{Berthier}, {Vachier}, {Thuillot}, {Fernique},
  {Ochsenbein}, {Genova}, {Lainey}  \& {Arlot}}{{Berthier}
  et~al.}{2006}]{Berthier2006}
{Berthier} J.,  {Vachier} F.,  {Thuillot} W.,  {Fernique} P.,  {Ochsenbein} F.,
   {Genova} F.,  {Lainey} V.,   {Arlot} J.-E.,  2006, in {Gabriel} C.,
  {Arviset} C.,  {Ponz} D.,   {Enrique} S.,  eds,  Astronomical Society of the
  Pacific Conference Series Vol. 351, Astronomical Data Analysis Software and
  Systems XV. pp 367--+

\bibitem[\protect\citeauthoryear{{Blagorodnova}, {Koposov}, {Wyrzykowski},
  {Irwin}  \& {Walton}}{{Blagorodnova} et~al.}{2014}]{Blagorodnova2014}
{Blagorodnova} N.,  {Koposov} S.~E.,  {Wyrzykowski} {\L}.,  {Irwin} M.,
  {Walton} N.~A.,  2014, \mn@doi [\mnras] {10.1093/mnras/stu837}, \href
  {http://adsabs.harvard.edu/abs/2014MNRAS.442..327B} {442, 327}

\bibitem[\protect\citeauthoryear{{Carrasco} et~al.,}{{Carrasco}
  et~al.}{2016}]{Carrasco2016}
{Carrasco} J.~M.,  et~al., 2016, \mn@doi [\aap] {10.1051/0004-6361/201629235},
  \href {http://adsabs.harvard.edu/abs/2016A%26A...595A...7C} {595, A7}

\bibitem[\protect\citeauthoryear{{Chambers} et~al.,}{{Chambers}
  et~al.}{2016}]{Chambers2016}
{Chambers} K.~C.,  et~al., 2016, preprint, \href
  {http://adsabs.harvard.edu/abs/2016arXiv161205560C} {} (\mn@eprint {arXiv}
  {1612.05560})

\bibitem[\protect\citeauthoryear{{Champion} et~al.,}{{Champion}
  et~al.}{2016}]{Champion2016}
{Champion} D.~J.,  et~al., 2016, \mn@doi [\mnras] {10.1093/mnrasl/slw069},
  \href {http://adsabs.harvard.edu/abs/2016MNRAS.460L..30C} {460, L30}

\bibitem[\protect\citeauthoryear{{Chatterjee} et~al.,}{{Chatterjee}
  et~al.}{2017}]{Chatterjee2017}
{Chatterjee} S.,  et~al., 2017, \mn@doi [\nat] {10.1038/nature20797}, \href
  {http://adsabs.harvard.edu/abs/2017Natur.541...58C} {541, 58}

\bibitem[\protect\citeauthoryear{{Clementini} et~al.,}{{Clementini}
  et~al.}{2016}]{Clementini2016}
{Clementini} G.,  et~al., 2016, \mn@doi [\aap] {10.1051/0004-6361/201629583},
  \href {http://adsabs.harvard.edu/abs/2016A%26A...595A.133C} {595, A133}

\bibitem[\protect\citeauthoryear{{Evans} et~al.,}{{Evans}
  et~al.}{2017}]{Evans2017}
{Evans} D.~W.,  et~al., 2017, \mn@doi [\aap] {10.1051/0004-6361/201629241},
  \href {http://adsabs.harvard.edu/abs/2017A%26A...600A..51E} {600, A51}

\bibitem[\protect\citeauthoryear{{Eyer} et~al.,}{{Eyer}
  et~al.}{2017}]{Eyer2017}
{Eyer} L.,  et~al., 2017, preprint, \href
  {http://adsabs.harvard.edu/abs/2017arXiv170203295E} {} (\mn@eprint {arXiv}
  {1702.03295})

\bibitem[\protect\citeauthoryear{{Fabricius} et~al.,}{{Fabricius}
  et~al.}{2016}]{Fabricius2016}
{Fabricius} C.,  et~al., 2016, \mn@doi [\aap] {10.1051/0004-6361/201628643},
  \href {http://adsabs.harvard.edu/abs/2016A%26A...595A...3F} {595, A3}

\bibitem[\protect\citeauthoryear{{Gaia Collaboration} et~al.,}{{Gaia
  Collaboration} et~al.}{2016a}]{Prusti2016}
{Gaia Collaboration} et~al., 2016a, \mn@doi [\aap]
  {10.1051/0004-6361/201629272}, \href
  {http://adsabs.harvard.edu/abs/2016A%26A...595A...1G} {595, A1}

\bibitem[\protect\citeauthoryear{{Gaia Collaboration} et~al.,}{{Gaia
  Collaboration} et~al.}{2016b}]{Brown2016}
{Gaia Collaboration} et~al., 2016b, \mn@doi [\aap]
  {10.1051/0004-6361/201629512}, \href
  {http://adsabs.harvard.edu/abs/2016A%26A...595A...2G} {595, A2}

\bibitem[\protect\citeauthoryear{{Gilliland} et~al.,}{{Gilliland}
  et~al.}{2010}]{Gilliland2010}
{Gilliland} R.~L.,  et~al., 2010, \mn@doi [\apjl]
  {10.1088/2041-8205/713/2/L160}, \href
  {http://adsabs.harvard.edu/abs/2010ApJ...713L.160G} {713, L160}

\bibitem[\protect\citeauthoryear{{Glennie}, {Jonker}, {Fender}, {Nagayama}  \&
  {Pretorius}}{{Glennie} et~al.}{2015}]{Glennie2015}
{Glennie} A.,  {Jonker} P.~G.,  {Fender} R.~P.,  {Nagayama} T.,   {Pretorius}
  M.~L.,  2015, \mn@doi [\mnras] {10.1093/mnras/stv801}, \href
  {http://adsabs.harvard.edu/abs/2015MNRAS.450.3765G} {450, 3765}

\bibitem[\protect\citeauthoryear{{G{\'o}rski}, {Hivon}, {Banday}, {Wandelt},
  {Hansen}, {Reinecke}  \& {Bartelmann}}{{G{\'o}rski}
  et~al.}{2005}]{Gorski2005}
{G{\'o}rski} K.~M.,  {Hivon} E.,  {Banday} A.~J.,  {Wandelt} B.~D.,  {Hansen}
  F.~K.,  {Reinecke} M.,   {Bartelmann} M.,  2005, \mn@doi [\apj]
  {10.1086/427976}, \href {http://adsabs.harvard.edu/abs/2005ApJ...622..759G}
  {622, 759}

\bibitem[\protect\citeauthoryear{{Hawley} et~al.,}{{Hawley}
  et~al.}{2002}]{Hawley2002}
{Hawley} S.~L.,  et~al., 2002, \mn@doi [\aj] {10.1086/340697}, \href
  {http://adsabs.harvard.edu/abs/2002AJ....123.3409H} {123, 3409}

\bibitem[\protect\citeauthoryear{{Hodgkin}, {Wyrzykowski}, {Blagorodnova}  \&
  {Koposov}}{{Hodgkin} et~al.}{2013}]{Hodgkin2013}
{Hodgkin} S.~T.,  {Wyrzykowski} L.,  {Blagorodnova} N.,   {Koposov} S.,  2013,
  \mn@doi [Philosophical Transactions of the Royal Society of London Series A]
  {10.1098/rsta.2012.0239}, \href
  {http://adsabs.harvard.edu/abs/2013RSPTA.37120239H} {371, 20120239}

\bibitem[\protect\citeauthoryear{{Jonker} et~al.,}{{Jonker}
  et~al.}{2013}]{Jonker2013}
{Jonker} P.~G.,  et~al., 2013, \mn@doi [\apj] {10.1088/0004-637X/779/1/14},
  \href {http://adsabs.harvard.edu/abs/2013ApJ...779...14J} {779, 14}

\bibitem[\protect\citeauthoryear{{LSST Science Collaboration} et~al.,}{{LSST
  Science Collaboration} et~al.}{2009}]{lsstsciencebook}
{LSST Science Collaboration} et~al., 2009, preprint, \href
  {http://adsabs.harvard.edu/abs/2009arXiv0912.0201L} {} (\mn@eprint {arXiv}
  {0912.0201})

\bibitem[\protect\citeauthoryear{{Lindegren} et~al.,}{{Lindegren}
  et~al.}{2016}]{Lindegren2016}
{Lindegren} L.,  et~al., 2016, \mn@doi [\aap] {10.1051/0004-6361/201628714},
  \href {http://adsabs.harvard.edu/abs/2016A%26A...595A...4L} {595, A4}

\bibitem[\protect\citeauthoryear{{Lorimer}, {Bailes}, {McLaughlin}, {Narkevic}
  \& {Crawford}}{{Lorimer} et~al.}{2007}]{Lorimer2007}
{Lorimer} D.~R.,  {Bailes} M.,  {McLaughlin} M.~A.,  {Narkevic} D.~J.,
  {Crawford} F.,  2007, \mn@doi [Science] {10.1126/science.1147532}, \href
  {http://adsabs.harvard.edu/abs/2007Sci...318..777L} {318, 777}

\bibitem[\protect\citeauthoryear{{MacLeod}, {Guillochon}, {Ramirez-Ruiz},
  {Kasen}  \& {Rosswog}}{{MacLeod} et~al.}{2016}]{Macleod2016}
{MacLeod} M.,  {Guillochon} J.,  {Ramirez-Ruiz} E.,  {Kasen} D.,   {Rosswog}
  S.,  2016, \mn@doi [\apj] {10.3847/0004-637X/819/1/3}, \href
  {http://adsabs.harvard.edu/abs/2016ApJ...819....3M} {819, 3}

\bibitem[\protect\citeauthoryear{{Macfarlane}, {Toma}, {Ramsay}, {Groot},
  {Woudt}, {Drew}, {Barentsen}  \& {Eisl{\"o}ffel}}{{Macfarlane}
  et~al.}{2015}]{Macfarlane2015}
{Macfarlane} S.~A.,  {Toma} R.,  {Ramsay} G.,  {Groot} P.~J.,  {Woudt} P.~A.,
  {Drew} J.~E.,  {Barentsen} G.,   {Eisl{\"o}ffel} J.,  2015, \mn@doi [\mnras]
  {10.1093/mnras/stv1989}, \href
  {http://adsabs.harvard.edu/abs/2015MNRAS.454..507M} {454, 507}

\bibitem[\protect\citeauthoryear{{Maehara}, {Shibayama}, {Notsu}, {Notsu},
  {Honda}, {Nogami}  \& {Shibata}}{{Maehara} et~al.}{2015}]{Maehara2015}
{Maehara} H.,  {Shibayama} T.,  {Notsu} Y.,  {Notsu} S.,  {Honda} S.,  {Nogami}
  D.,   {Shibata} K.,  2015, \mn@doi [Earth, Planets, and Space]
  {10.1186/s40623-015-0217-z}, \href
  {http://adsabs.harvard.edu/abs/2015EP%26S...67...59M} {67, 59}

\bibitem[\protect\citeauthoryear{{Pedersen} et~al.,}{{Pedersen}
  et~al.}{1984}]{Pedersen1984}
{Pedersen} H.,  et~al., 1984, \mn@doi [\nat] {10.1038/312046a0}, \href
  {http://adsabs.harvard.edu/abs/1984Natur.312...46P} {312, 46}

\bibitem[\protect\citeauthoryear{{Price-Whelan} et~al.,}{{Price-Whelan}
  et~al.}{2014}]{Pricewhelan2014}
{Price-Whelan} A.~M.,  et~al., 2014, \mn@doi [\apj]
  {10.1088/0004-637X/781/1/35}, \href
  {http://adsabs.harvard.edu/abs/2014ApJ...781...35P} {781, 35}

\bibitem[\protect\citeauthoryear{{Roelens} et~al.,}{{Roelens}
  et~al.}{2017}]{Roelens2017}
{Roelens} M.,  et~al., 2017, preprint, \href
  {http://adsabs.harvard.edu/abs/2017arXiv170808703R} {} (\mn@eprint {arXiv}
  {1708.08703})

\bibitem[\protect\citeauthoryear{{Rosswog}, {Ramirez-Ruiz}  \& {Hix}}{{Rosswog}
  et~al.}{2009}]{Rosswog2009}
{Rosswog} S.,  {Ramirez-Ruiz} E.,   {Hix} W.~R.,  2009, \mn@doi [\apj]
  {10.1088/0004-637X/695/1/404}, \href
  {http://adsabs.harvard.edu/abs/2009ApJ...695..404R} {695, 404}

\bibitem[\protect\citeauthoryear{{Skrutskie} et~al.,}{{Skrutskie}
  et~al.}{2006}]{Skrutskie2006}
{Skrutskie} M.~F.,  et~al., 2006, \mn@doi [\aj] {10.1086/498708}, \href
  {http://adsabs.harvard.edu/abs/2006AJ....131.1163S} {131, 1163}

\bibitem[\protect\citeauthoryear{{Spitler} et~al.,}{{Spitler}
  et~al.}{2016}]{Spitler2016}
{Spitler} L.~G.,  et~al., 2016, \mn@doi [\nat] {10.1038/nature17168}, \href
  {http://adsabs.harvard.edu/abs/2016Natur.531..202S} {531, 202}

\bibitem[\protect\citeauthoryear{{Wright} et~al.,}{{Wright}
  et~al.}{2010}]{wright2010}
{Wright} E.~L.,  et~al., 2010, \mn@doi [\aj] {10.1088/0004-6256/140/6/1868},
  \href {http://adsabs.harvard.edu/abs/2010AJ....140.1868W} {140, 1868}

\bibitem[\protect\citeauthoryear{{Wyrzykowski} et~al.,}{{Wyrzykowski}
  et~al.}{2016}]{Wyrzykowski2016}
{Wyrzykowski} {\L}.,  et~al., 2016, \mn@doi [\mnras] {10.1093/mnras/stw426},
  \href {http://adsabs.harvard.edu/abs/2016MNRAS.458.3012W} {458, 3012}

\bibitem[\protect\citeauthoryear{{Yi}, {Yu}, {Wang}  \& {Dai}}{{Yi}
  et~al.}{2017}]{Yi2017}
{Yi} S.-X.,  {Yu} H.,  {Wang} F.~Y.,   {Dai} Z.~G.,  2017, preprint, \href
  {http://adsabs.harvard.edu/abs/2017arXiv170608716Y} {} (\mn@eprint {arXiv}
  {1706.08716})

\bibitem[\protect\citeauthoryear{{van Leeuwen} et~al.,}{{van Leeuwen}
  et~al.}{2017}]{Leeuwen2017}
{van Leeuwen} F.,  et~al., 2017, \mn@doi [\aap] {10.1051/0004-6361/201630064},
  \href {http://adsabs.harvard.edu/abs/2017A%26A...599A..32V} {599, A32}

\bibitem[\protect\citeauthoryear{von Neumann}{von Neumann}{1941}]{Neumann1941}
von Neumann J.,  1941, \mn@doi [Ann. Math. Statist.] {10.1214/aoms/1177731677},
  12, 367

\makeatother
\end{thebibliography}

\section*{Appendix: photometric data of the outlying transits of the 4 strong candidates}
\FloatBarrier
\begin{table}
 \centering
  \caption{Photometric data of the outlying transit in FGT1. The magnitude is quoted in the \gaia $G$-band.}
  \begin{tabular}{ccc}
  \hline
  JD - 2456000 & Magnitude & Error\\
  \hline
1791.86581 & 18.25 & 0.01\\
1791.86703 & 18.26 & 0.01\\
1791.86826 & 18.26 & 0.01\\
1791.86948 & 18.34 & 0.01\\
1791.87070 & 18.30 & 0.01\\
1791.87192 & 18.33 & 0.01\\
1791.87315 & 18.36 & 0.01\\
1791.87437 & 18.35 & 0.01\\
1791.87559 & 18.40 & 0.01\\
1791.87681 & 18.41 & 0.01\\
\hline
    \end{tabular}
  \label{tab:fgt1lc}
\end{table}
\begin{table}
 \centering
  \caption{Photometric data of the outlying transit in FGT2. The magnitude is quoted in the \gaia $G$-band.}
  \begin{tabular}{ccc}
  \hline
  JD - 2456000 & Magnitude & Error\\
  \hline
1409.19602 & 17.26 & 0.01\\
1409.19724 & 17.27 & 0.01\\
1409.19846 & 17.25 & 0.01\\
1409.19968 & 17.16 & 0.01\\
1409.20090 & 17.06 & 0.01\\
1409.20213 & 17.02 & 0.01\\
1409.20335 & 17.02 & 0.01\\
1409.20457 & 17.01 & 0.01\\
1409.20579 & 17.16 & 0.01\\
1409.20702 & 17.25 & 0.01\\
\hline
    \end{tabular}
  \label{tab:fgt2lc}
\end{table}

\begin{table}
 \centering
  \caption{Photometric data of the outlying transit in FGT3. The magnitude is quoted in the \gaia $G$-band.}
  \begin{tabular}{ccc}
  \hline
  JD - 2456000 & Magnitude & Error\\
  \hline
1313.78254 & 19.03 & 0.02\\
1313.78376 & 19.04 & 0.02\\
1313.78499 & 19.06 & 0.02\\
1313.78621 & 18.97 & 0.02\\
1313.78743 & 18.95 & 0.02\\
1313.78865 & 18.84 & 0.02\\
1313.78987 & 18.56 & 0.02\\
1313.79110 & 18.21 & 0.02\\
1313.79232 & 17.97 & 0.01\\
1313.79354 & 17.81 & 0.01\\
\hline
    \end{tabular}
  \label{tab:fgt3lc}
\end{table}

\begin{table}
 \centering
  \caption{Photometric data of the outlying transit in FGT4. The magnitude is quoted in the \gaia $G$-band.}
  \begin{tabular}{ccc}
  \hline
  JD - 2456000 & Magnitude & Error\\
  \hline
1660.64121 & 18.31 & 0.01\\
1660.64243 & 18.40 & 0.01\\
1660.64365 & 18.31 & 0.01\\
1660.64488 & 18.31 & 0.01\\
1660.64610 & 18.35 & 0.01\\
1660.64732 & 18.32 & 0.01\\
1660.64854 & 18.41 & 0.01\\
1660.64977 & 18.39 & 0.01\\
1660.65099 & 18.36 & 0.01\\
1660.65221 & 18.38 & 0.01\\
\hline
    \end{tabular}
  \label{tab:fgt4lc}
\end{table}

\label{lastpage}
\end{document}